\documentclass{PoS}

\usepackage{amssymb}
\newcommand{\bc}{\begin{center}}
\newcommand{\ec}{\end{center}}
\newcommand{\be}{\begin{equation}}
\newcommand{\ee}{\end{equation}}

\newcommand{\tr}{\hbox{tr}}

\title{Glueballs and mesons in the superfluid phase of two color QCD}

\ShortTitle{Glueballs and dense Q$C_2$D}

\author {\speaker{Maria Paola~Lombardo}\\
Istituto Nazionale di Fisica Nucleare, LNF, Via Enrico Fermi 40, I 00044, Frascati (Roma), Italy\\
E-mail: \email{lombardo@lnf.infn.it}}

\author{Maria Luigia Paciello\\
INFN, Sezione di  Roma,
P.le A. Moro 2, I-00185 Roma, Italy\\
E-mail:\email{mariella.paciello@roma1.infn.it}}

\author{Silvano Petrarca \\Dipartimento di Fisica, Universit\`a di Roma ``La
Sapienza'', \\P.le A. Moro 2, I-00185 Roma, Italy, and INFN, Sezione di Roma\\
E-mail:\email{silvano.petrarca@roma1.infn.it}}

\author{Bruno Taglienti\\INFN, Sezione di  Roma,
P.le A. Moro 2, I-00185 Roma, Italy\\
E-mail:\email{bruno.taglienti@roma1.infn.it}}

\abstract{QCD with two colors undergoes a transition to a 
superfluid phase with diquark
condensate when the quark chemical potential equals half the pion mass.
We  investigate the gluonic aspects of the  transition
by inspecting the behavior of the glueball correlators evaluated via a
multi-step smearing procedure for several values of chemical potential 
ranging between zero and the saturation
threshold.
The results are based on an analysis of $0^{++}$ glueball correlators, 
on a sample of  40000 independent configurations on each parameter
set.  The amplitudes of the correlators peak for $\mu = m_\pi/2$,
 indicating that the superfluid phase transition affects the
gluonic sector as well. The mass of the fundamental state decreases 
in the superfluid phase, and the amplitude of the propagators drops, 
suggesting a reduction of the gluon condensate, in agreement with model
calculations.  The analysis of the smearing dependence
of the results helps disentangling the role of long and short distance
phenomena at the superfluid transition.}

\FullConference{The XXV International Symposium on Lattice Field Theory\\
		 July 30-4 August 2007\\
		 Regensburg, Germany}

\begin{document}

\section{Introduction, and simulation setup}
\label{intro}

QCD-like models whose determinant remains real at nonzero chemical potential
afford the possibility of standard Monte Carlo simulations.
Two-color QCD is one such model, which has been extensively studied over
the past few years. Many results have been obtained in the fermionic sector,
 see e.g. \cite{Sinclair:2003rm} for a 
concise review, while gluodynamics is comparably much less known. 
 However, gluodynamics is arguably the sector 
where results from two-color QCD 
have most direct relevance to real QCD. The few 
studies of finite density gluodynamics performed so far concentrated 
on the Polyakov loop \cite{gise}, 
topological susceptibility \cite{Alles:2006ea} and gluon propagator
\cite{Hands:2006ve}. In this note - which continues
a work initiated in ref. \cite{tsukuba} -- we study
glueball correlators at finite density. A more complete account of
our results, with a complete set of references,  will appear soon \cite{tbs}.
In the same paper we will present, together with the glueball spectrum,
a high statistics update of the results on the meson spectrum presented
in \cite{tsukuba}. Since the new results in the fermionic sector 
merely confirm our previous findings, we will concentrate here on the 
glueball results.

The simulations were performed using a standard
hybrid molecular dynamics algorithm for eight continuum
flavors of staggered fermions \cite{hklm}
on a $6^3 \times 12$ lattice. We accumulated 40000 MC trajectories of
unit length for each value of the chemical potential and two quark masses,
m=0.05 and m=0.07. 

We have measured the fermionic 
spectrum at $\mu=0.0$ , obtaining $m_\pi = 0.56 (2)$ and $m_\rho = 1.4(1)$
for $m=0.05$ , and $m_\rho = 1.5(2)$ and $m_\pi$ = 0,64 for m=0.07. 
In both cases $m_\rho$ and $m_\pi$ are well resolved, with the
superfluid transition at $\mu = m_\pi/$ well below the threshold of the 
predicted
vector condensation , $\mu = m_\rho/2$\cite{San}.
Moreover, $m_\pi/m_\rho < 1.$ implies
that we are still in the region where chiral perturbation theory
can be safely applied. If, just to get an idea of the lattice spacing,
 we extrapolate linearly $m_\rho$ according to $m_\rho(m) = m_\rho(0) -k m$  
we arrive at $m_\rho(.0) = 1.1$ in lattice units.

\section{Glueball measurements: operators and smearing}
\label{op}

The operators commonly used for measuring scalar gluonic correlators exciting 
glueball mass are Wilson loops. For simplicity we restricted ourselves to 
plaquette-like operators that can be built from four links. 

Simple glueball wave functions such as the plaquette have 
small overlaps with the lowest-lying glueball 
states. Moreover, the overlaps become rapidly smaller as the lattice spacing 
is decreased. Furthermore  the plaquette couples strongly to ultraviolet 
fluctuations, increasing the noise in the correlators.
To have reliable glueball correlation functions at different distances, 
it is mandatory to reduce statistical fluctuations. 

Different smoothing procedure and methods to remove 
the unphysical short-distance fluctuations have been introduced.
A short review can be found in \cite{review}.
 
One of the most successful approach is the smearing method. 
This procedure, originally 
proposed for pure gauge SU(3) in \cite{smearing},  
consists in the construction of 
correlation functions of operators which are 
a functional of the field smeared in space and not in time. 

This procedure reduces the noise associated with lattice artifacts and
 can be 
iterated.  Only spatial links participate on the averaging. Thus the transfer
 matrix  is not affected by the smearing procedure and remains positive 
definite. 
The values of the smearing coefficient and the iterations are  tuned in order 
to optimize the performance of the method \cite{alba,calibro,gupta}. 

We analyze glueball states with zero momentum in the $A1^{++}$ 
(one-dimensional) irreducible representations of the  relevant 
cubic point 
group  (on a lattice the full rotational symmetry is broken down to only 
cubic symmetry). The  $E^{++}$ (2-dimensional) turns out to be, as expected,
extremely noisy and in the present note we do not discuss it.

The $A_{1}^{++}$ and $E^{++}$  
representations correspond
to the continuum $O(3) \otimes Z(2)$, $J^{PC}$ = $0^{++}$ and $2^{++}$ 
respectively. We can then  label
the associated glueballs as $0^{++}$ and  $2^{++}$ states.  Of course this 
correspondence is not one to one but infinite to one. Therefore what we can 
measure is the lowest excitation in the corresponding representation of the 
cubic group. With dynamical fermions mixing is possible with fermionic
states, so strictly speaking we should always use the expression 'lowest
excitation in the corresponding representation  of the cubic group'
rather then 'gluebal mass'.

The glueball operators are defined by means of the plaquettes 
$P_{ij}(\vec{x},t)$ on $ij$ plane as in the following:
\begin{equation}
\phi ^{0^{++}} (t) = \tr \sum _{\vec{x} } \biggl[ 
P_{12}(\vec{x},t) \, + \, P_{23}(\vec{x},t) \, + \, 
P_{13}(\vec{x},t) \biggr] 
\label{eq:oper0}
\end{equation}
which transforms according to the $A_{1}^{++}$ representation and couples 
to the scalar glueball $0^{++}$, and
\begin{eqnarray}
\phi_a ^{2^{++}} (t) &=& \tr \sum _{\vec{x} } \biggl[ 
P_{12}(\vec{x},t) \, - \, P_{13}(\vec{x},t) \biggr]
\label{eq:oper2a}\\
 \phi_b ^{2^{++}} (t) &=& \tr \sum _{\vec{x} } \biggl[ 
P_{12}(\vec{x},t) \, + \, P_{23}(\vec{x},t) \, - \, 
2 P_{13}(\vec{x},t) \biggr] 
\label{eq:oper2b} 
\end{eqnarray}  
which transform both according to the $E^{++}$ representations and couples 
to the tensor glueball $2^{++}$.

Higher levels of smearing are obtained by varying the weight $w$ and by 
iterating the procedure $N_s$ times.

Glueball masses are calculated from the behavior of the correlation functions.
We have analyzed  glueball correlation as function of smearing parameters. 
In effect, in ref. \cite{calibro}, it has been shown that to a  a good 
approximation the two-dimensional parameters space of the number of 
sweeps $N_s$ and the smearing weight $w$  may be reduced  to a single 
dimension  via the parameter $T_s=N_s\times w$.

To assess the optimal parameter choice in our study, 
we have analyzed in some more detail the deviations from the 
unidimensional parametrization. The results of Figs. \ref{Fig:unid} below 
show that the unidimensional parametrization remains true till 
$w \le 0.3$, irrespective of the number of smearing steps, at least
within the allowed range of smearing steps $N_s \le S/2$, where $S$
is the spatial size of the lattice. 

To be on the safe side, we will base our 
discussions on results within the range of validity of the unidimensional
parametrization,taking into account the limitations imposed
by the spatial size of the lattice: in conclusion, $w \le 0.3, N_s \le 3$.

\begin{figure}[t]
\centering
\vspace*{-110mm}
\begin{minipage}[t]{0.45\textwidth}
\hspace*{-30mm}
\includegraphics[width=1.9\textwidth]{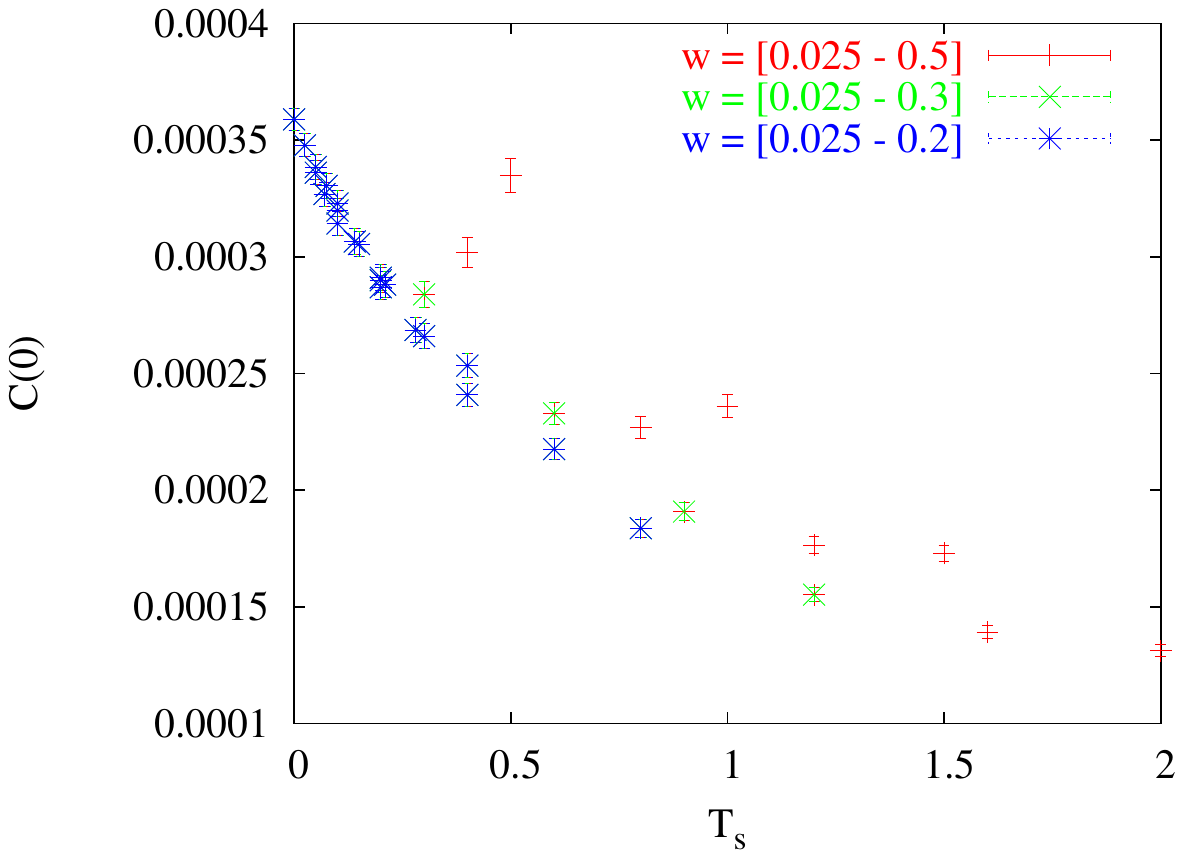}
\end{minipage}
\hspace*{-20mm}
\begin{minipage}[t]{0.45\textwidth}
\includegraphics[width=1.9\textwidth]{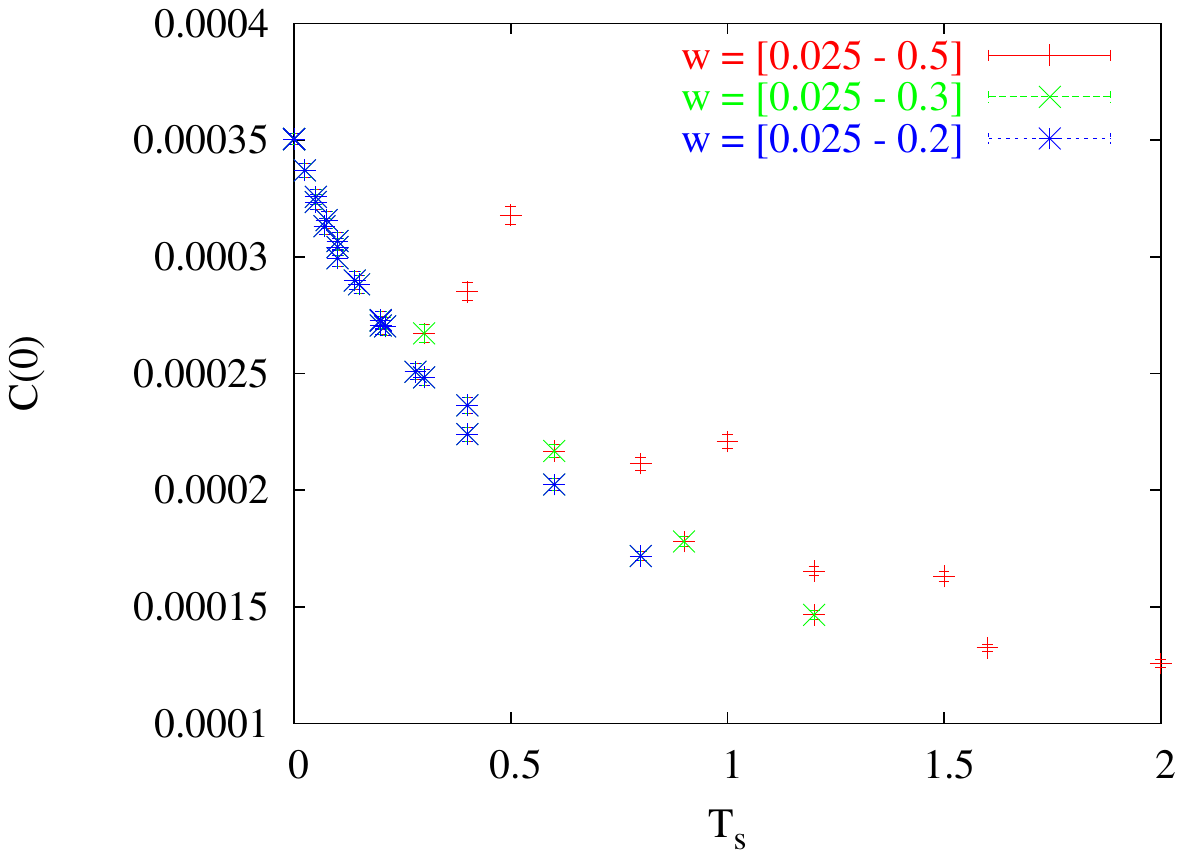}
\end{minipage}

\vspace*{-110mm}
\begin{minipage}[t]{0.45\textwidth}
\hspace*{-30mm}
\includegraphics[width=1.9\textwidth]{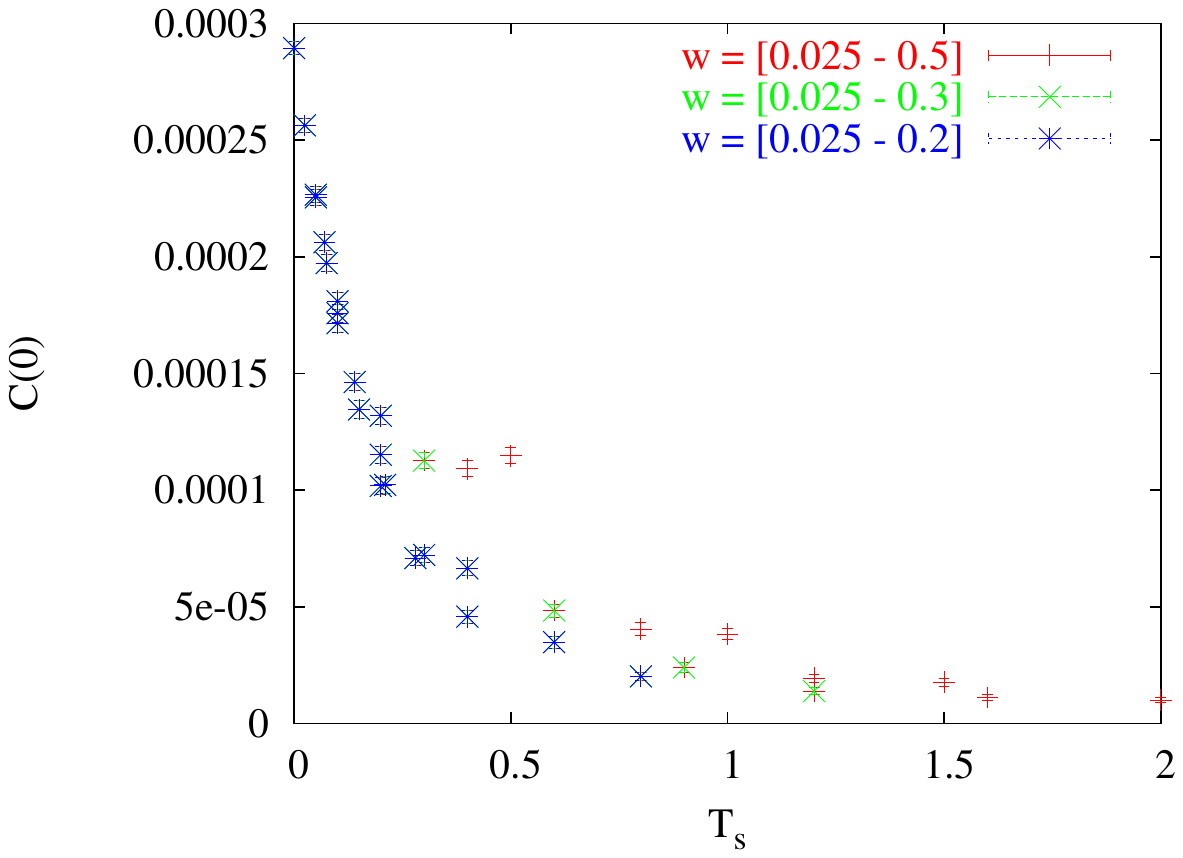}
\end{minipage}
\hspace*{-20mm}
\begin{minipage}[t]{0.45\textwidth}
\includegraphics[width=1.9\textwidth]{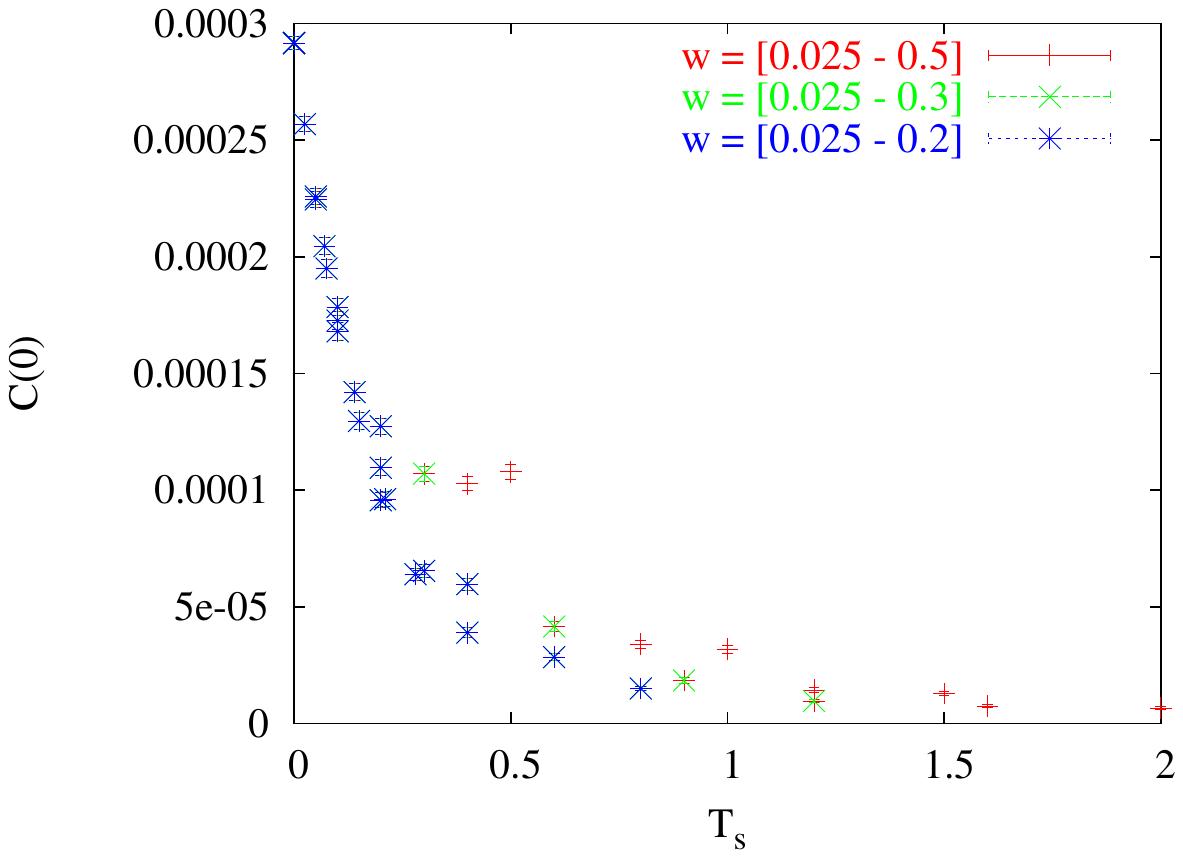}
\end{minipage}

\vspace*{-15mm}
\caption{Amplitude of the $0^{++}$ correlator
 as a function of the smearing parameter $T_s$, smeraring steps 1 to 4,
and smearing weights as indicated. 'Universality' holds till $w  \le 0.3$.
$\mu=0, m=0.05$ and 0.07 (upper); $\mu=0.6$ and m = 0.05 and 0.07, lower }
\label{Fig:unid}
\end{figure}

\section{Results}

\subsection{Amplitudes}

If we consider
the cluster properties of the correlator themselves, we  immediately 
associate the amplitudes with plaquette susceptibilities. A second
intepretation, albeit indirect, associate the amplitude with the gluon
condensate.

Interestingly, and not surprisingly! the amplitudes peak at the 
critical point $\mu_c = m_\pi/$, providing a very clean estimate of
the  position of the critical point itself, and a clear-cut evidence that
the critical behavious seen in the fermionic sector shows up in purely
gluonic observables as well.

At larger $\mu$, close to the saturation region, the amplitudes increase
again, catching up with the quenched results as they should
(see e.g. \cite{Alles:2006ea}).

Note that the amplitudes are ultraviolet divergent. However, the smearing
procedure\cite{smearing}-which we have reviewd above- 
should remove these divergencies, togheter with other short
distance artifacts. In the bona fide superfluid region 
- $m_\pi/2 < \mu < m_\rho/2$ our results show that the
effect of smearing is more significant that in the normal phase: this
further indicates the reduction of the soft, non-perturbative component
of the propagators in this phase. Note that the gluonic condensate contributes 
to the amplitude, hence the
reduction of the amplitudes in the superfluid phase is consistent with
the decrease of the gluon condensate predicted by model studies
\cite{Zhitnitsky:2007uk}.

\begin{figure}[t]
\centering
\vspace*{-110mm}
\begin{minipage}[t]{0.45\textwidth}
\hspace*{-30mm}
\includegraphics[width=1.9\textwidth]{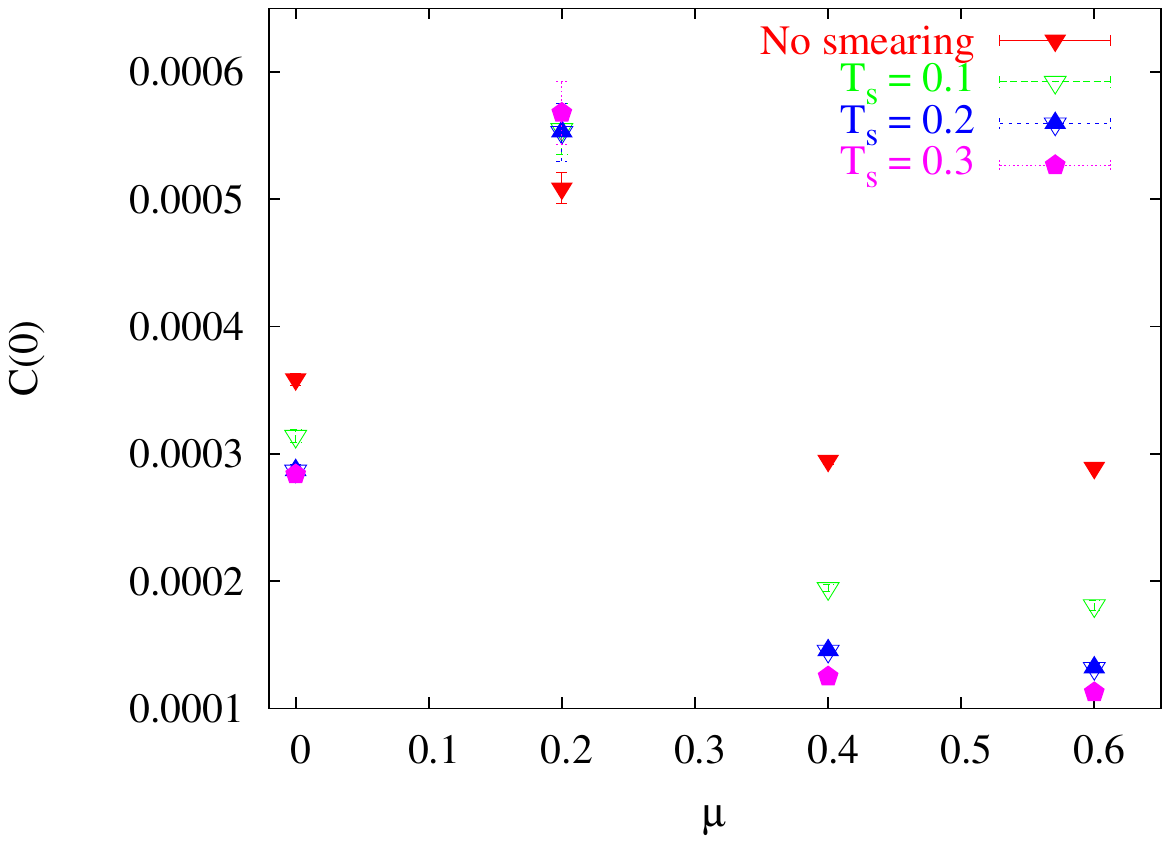}
\end{minipage}
\hspace*{-20mm}
\begin{minipage}[t]{0.45\textwidth}
\includegraphics[width=1.9\textwidth]{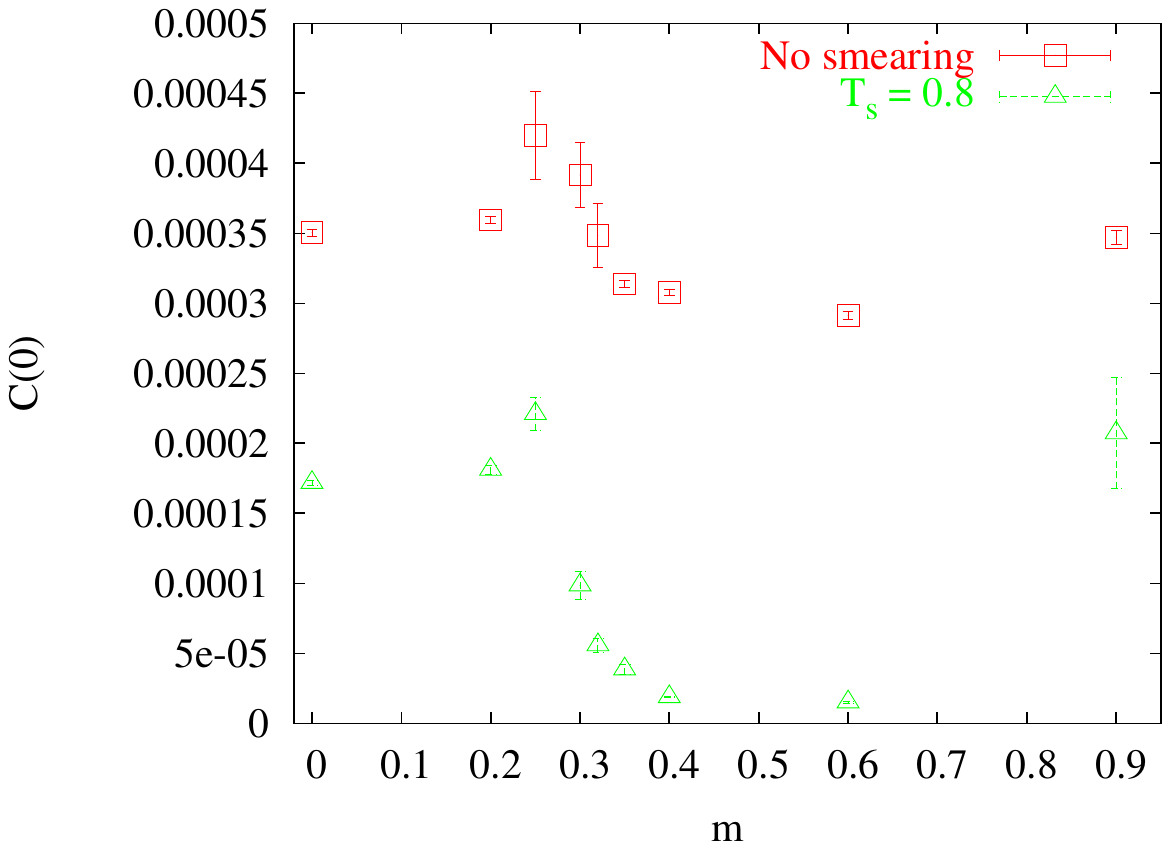}
\end{minipage}

\vspace*{-12mm}
\caption{Zero dinstance correlations as a function of $\mu$ for two values
of $T_s$ in the safe region, and quark mass = 0.05(left) and 0.07}
\label{fig:amplimu_m5}
\end{figure}

\subsection{The superfluid phase}

\begin{figure}[t]
\centering
\vspace*{-100mm}
\begin{minipage}[t]{0.45\textwidth}
\hspace*{-30mm}
\includegraphics[width=1.9\textwidth]{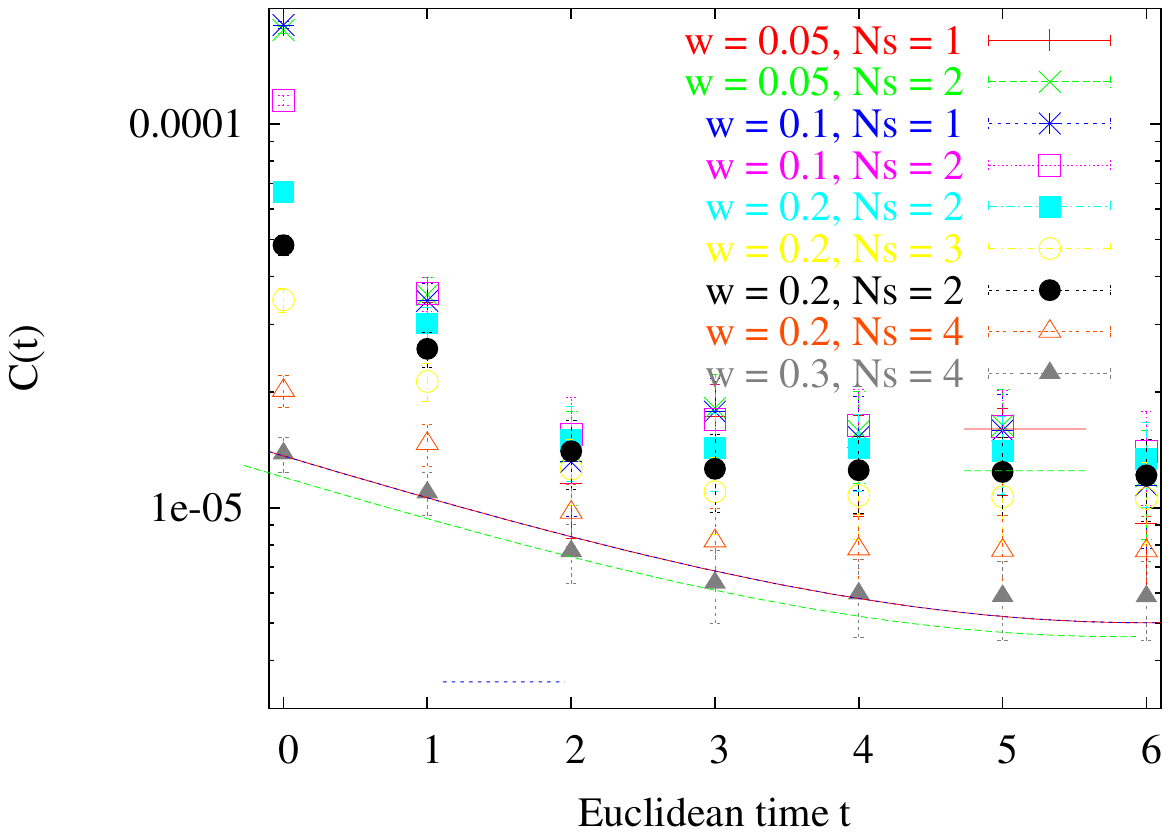}
\end{minipage}
\hspace*{-20mm}
\begin{minipage}[t]{0.45\textwidth}
\includegraphics[width=1.9\textwidth]{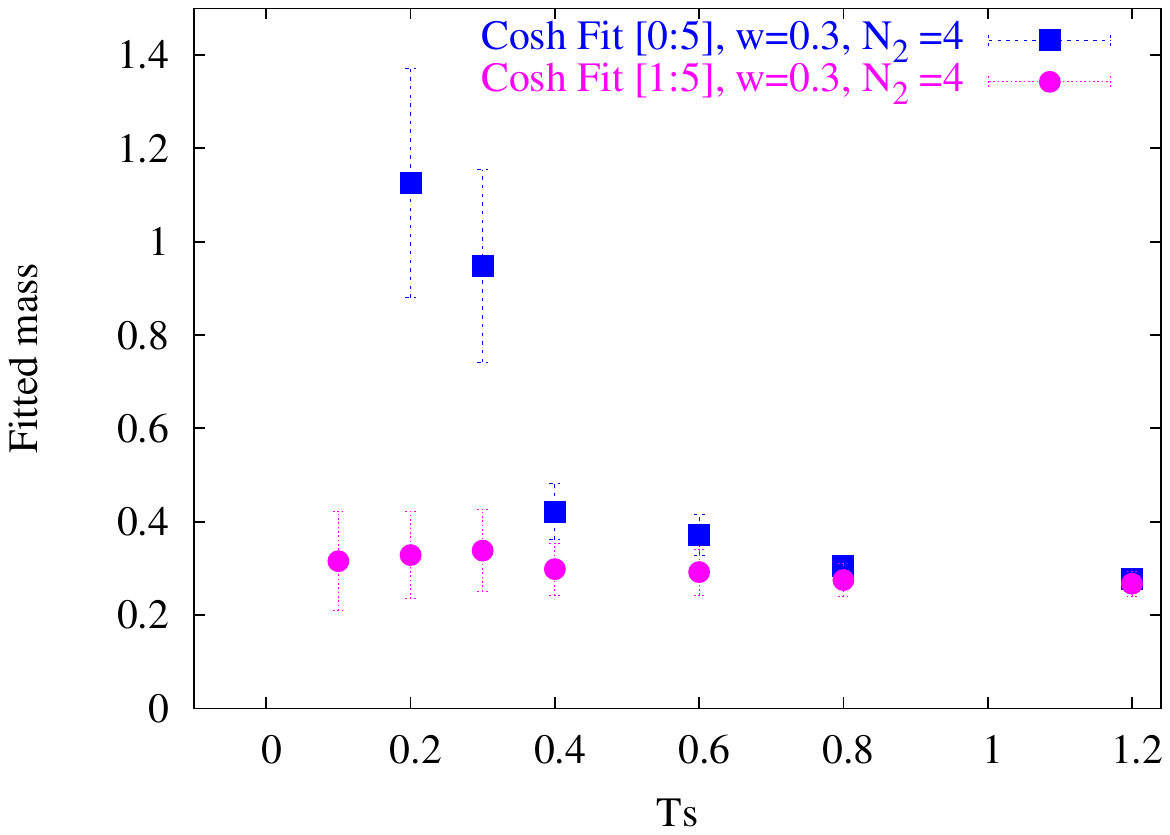}
\end{minipage}
\vspace*{-11mm}
\caption{Sample of fits(left) and fit results for two different time
intervals as a function of the parameter 
$T_s=N_s\times w$ for $\mu=0.6$  and $m=0.05$.
}
\label{Fig:fitres}
\end{figure}

The glueball propagators in the superfluid phase are amenable to
a standard analysis based on hyperbolic cosine fits, 
supporting the view that glueballs still exist
as bound states in this phase. The main observation - supported
either by the results of the fits -- see Figure \ref{Fig:fitres} --
by the effective mass analysis and by a direct comparisons with the normal
phase  (see next Section) is that the lighest excitations in the gluonic
channel in the superluid phase is ligher than in the normal phase.

The fits  show that the scalar glueball 
 looks a reasonable bound state in the superfluid phase since
it can be easily fitted to a simple  $\cosh (m t)$ 
form. There is no direct
indication of a modification of the associated spectral function,
and this holds true for both masses.
Aside, it is interesting to compare our observations with the results 
of a study of the glueball spectrum at nonzero temperature 
\cite{ishii1,ishii2}.

\begin{figure}[t]
\vspace*{-110mm}
\begin{minipage}[t]{0.45\textwidth}
\hspace*{-12mm}
\includegraphics[width=1.9\textwidth]{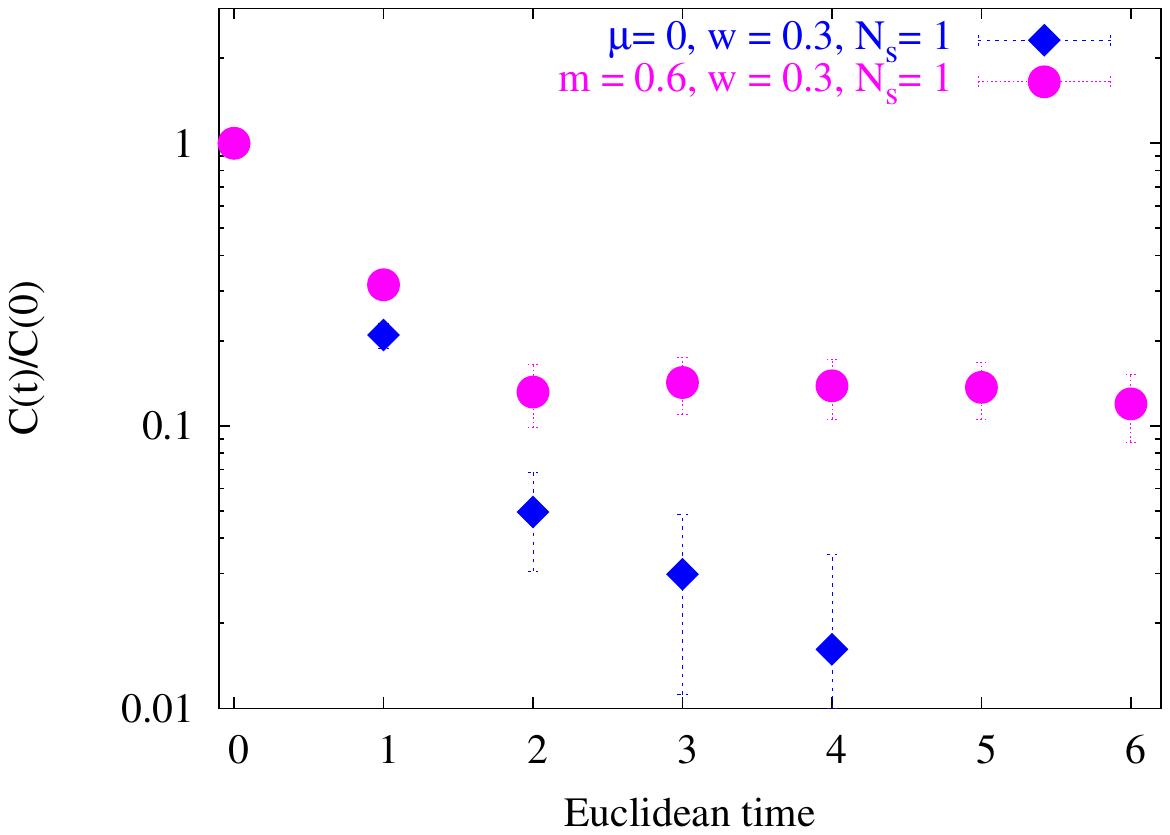}
\end{minipage}
\hspace*{-4mm}
\begin{minipage}[t]{0.45\textwidth}
\includegraphics[width=1.9\textwidth]{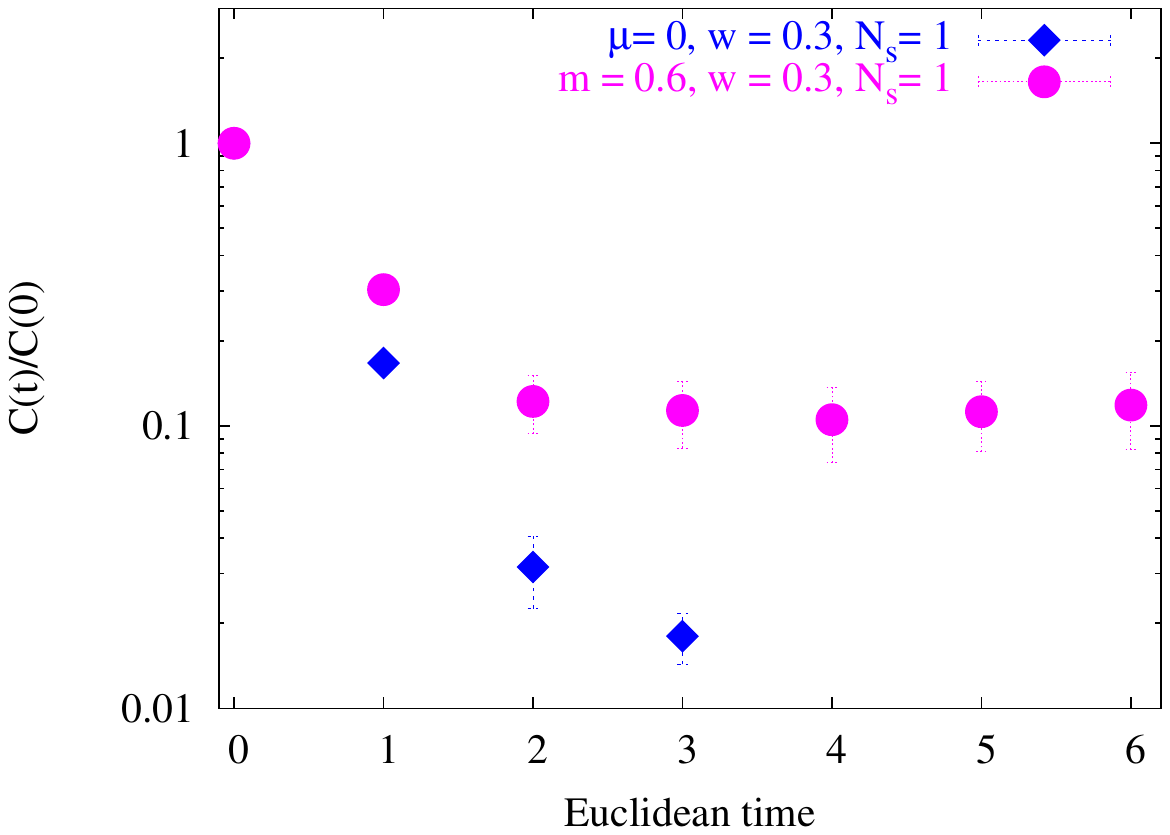}
\end{minipage}
\vspace*{-11mm}
\caption{Propagators of the scalar glueball at mass = 0.05(0.07) , left(right)
in the normal and superfluid phase, normalized to one at zero distance}
\label{Fig:compa}
\end{figure}

It is then meaningful
to directly compare the glueball correlators in the two phases,
which is done in Figure \ref{Fig:compa}. These results show that the
lowest excitation with $0^{++}$ quantum numbers becomes ligher in the
superfluid phase.

\subsection{The critical region}
\label{cri}

In the critical region we found an unexpected oscillatory behaviour
of the glueball propagator. To a much lesser extent, this can 
still be perceived  in the superfluid phase, although, as discussed above, 
at $\mu=0.6$
the results are again well represented by a conventional hyperbolic
cosine behaviour.

The behaviour was observed first at $m = 0.05, \mu = 0.2$. Subsequently
we did a fine $\mu$ scan of the critical region at $m= 0.07$ and 
found evidence for the same behaviour.

The conclusion from our numerical study is that the propagators
in the critical region  are well described by
\begin{equation}
C(t) = A e^{-mt}cos(\alpha t)
\end{equation}
In words, the propagators develop complex poles in the critical region, which
come in complex conjugate pairs.

Clearly, because of the lattice periodicity in time, the possible periods
are strictly quantized, and, as soon as a very small imaginary component
of the pole of the propagator develops, it manifest itself in a period 
of $N_t=12$  lattice units. Heavier poles might be observables
and give rise to smaller periods, which is probably what we observe
at large $\mu$. They should eventually decouple or disappear. Only
a scaling analysis either with $N_t$ and with the coupling can answer
this question.

\begin{figure}[t]
\vspace*{-100mm}
\begin{minipage}[t]{0.45\textwidth}
\hspace*{-12mm}
\includegraphics[width=1.9\textwidth]{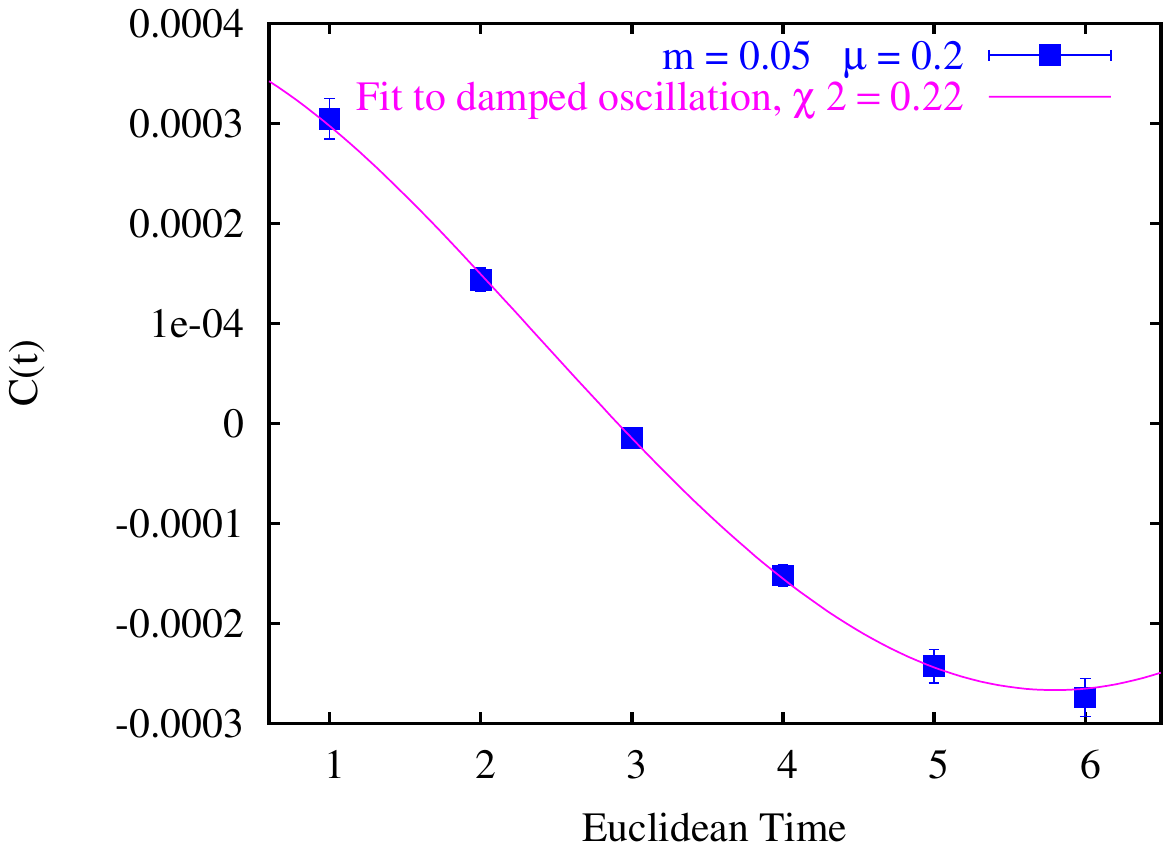}
\end{minipage}
\hspace*{-4mm}
\begin{minipage}[t]{0.45\textwidth}
\includegraphics[width=1.9\textwidth]{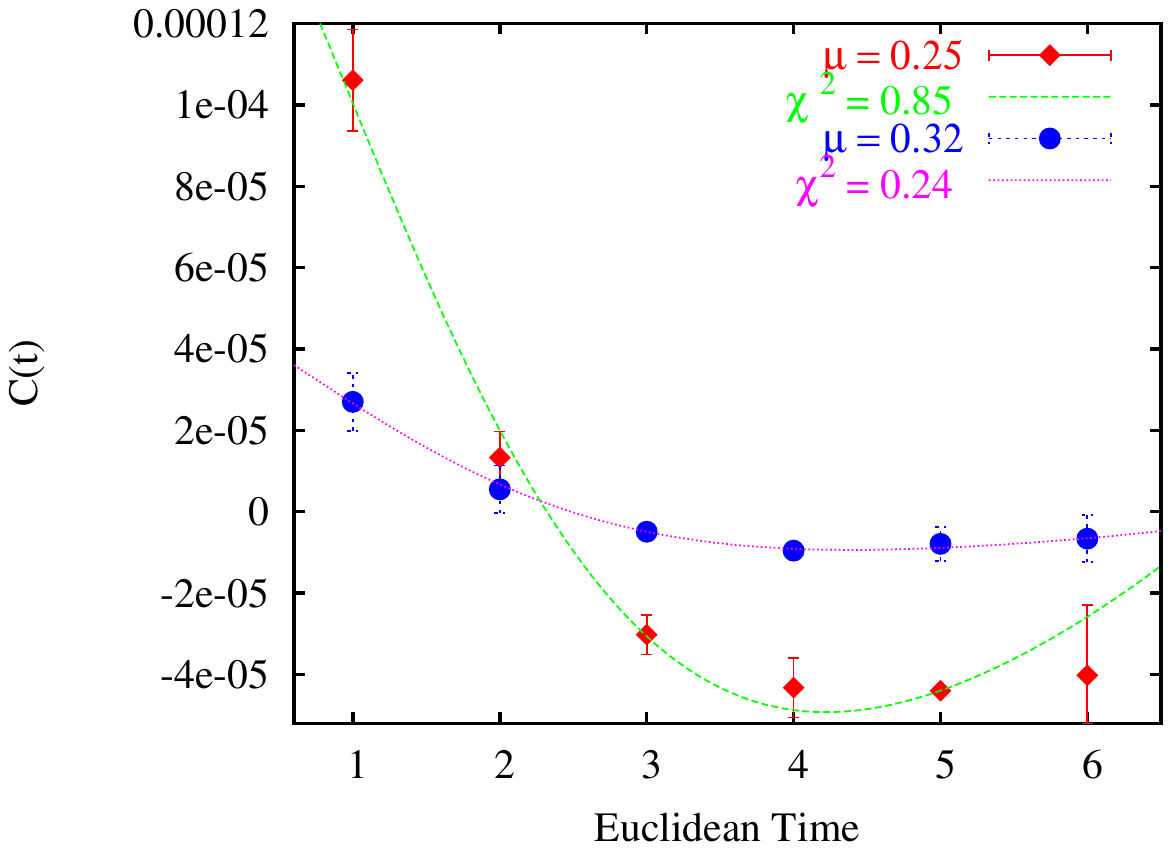}
\end{minipage}
\vspace*{-11mm}
\caption{The oscillatory behaviour at $m=.0.05$, $\mu = 0.2$ with
supermposed the fit described in the text (left). two representative
results at $m=0.07$ (right)}
\label{Fig:osci}
\end{figure}

\subsection{Summary}
\label{con}

We have observed a  gluonic transition coincident
with the superfluid transition at $\mu = m_\pi/2$,
for $m=0.05$ and $m=0.07$,
associated with a peak in the amplitude of the glueball propagator.

In the superfluid region the (smeared) amplitude of the
propagator is suppressed, signalling a reduction of the non--perturbative,
large distance contributions to the gauge dynamics - i.e.
a reduction of the gluon condensates.

Glueballs propagators in the superfluid phase, away from
the critical point, can be fitted by a conventional $\cosh (m t)$ 
behaviour. This, together with the observation that
the Polyakov loop apparently is not sensitive to this transition 
\cite{tbs} confirms that the superfluid phase is confining 
\cite{Hands:2006ve,Zhitnitsky:2007uk}. 
This should be contrasted
with observations at finite density, and a nonzero temperatures
where the transition is indeed deconfining\cite{Muroya:2002ry,gise,Alles:2006ea}

A direct comparison of the propagators in the two phases
shows that the lowest mode in the $0^{++}$ channel
is much lighter in the superfluid phase than in the normal phase. This
confirms and extends our previous results on the meson and glueball
level ordering \cite{tsukuba}.

In the critical region the glueball propagators are dominated
by a complex pole. This might indicate  some structure in the 4d space-time 
 possibly amplified by lattice artifacts. A study of 
the $N_t$ scaling, closer to the continuum limit,  should set this issue.

\end{document}